\documentclass[useAMS,usenatbib]{mn2e}
\usepackage{natbib}
\usepackage{graphicx}
\title[Velocity dispersion in MOND]{Testing fundamental physics with distant star clusters: theoretical models for pressure-supported  stellar systems}
\author[Haghi et al.]
{Hosein Haghi$^{1,2}$\thanks{E-mail: \mbox{haghi@iasbs.ac.ir} (HH);
\mbox{holger@astro.uni-bonn.de} (HB); \mbox{pavel@astro.uni-bonn.de}
(PK); \mbox{grebel@ari.uni-heidelberg.de} (EKG);
\mbox{mhilker@eso.org}(MH); \mbox{jordi@ari.uni-heidelberg.de}
(KJ)}, Holger Baumgardt$^2$, Pavel Kroupa$^2$, Eva K. Grebel$^{3}$,
\newauthor
Michael Hilker$^{4}$and Katrin Jordi$^{3}$\\
$^{1}$Department of Physics, Institute for Advanced Studies in Basic
Sciences (IASBS), P. O. Box 45195-1159, Zanjan, 45195, Iran\\
$^{2}$Argelander Institute for Astronomy (AIfA), Auf dem H\"ugel 71, D-53121 Bonn, Germany\\
$^{3}$Astronomisches Rechen-Institut, Zentrum fuer Astronomie, Universitaet Heidelberg/Germany\\
$^{4}$ESO, Garching/Germany }
\begin{document}
\date{Accepted \ldots. Received \ldots; in original form \ldots}
\pagerange{\pageref{firstpage}--\pageref{lastpage}} \pubyear{2008}
\maketitle

\label{firstpage}

\maketitle
\begin{abstract}
We investigate the mean velocity dispersion and the velocity
dispersion profile of stellar systems in MOND, using the N-body code
N-MODY, which is a particle-mesh based code with a numerical MOND
potential solver developed by Ciotti, Londrillo and Nipoti (2006).
We have calculated mean velocity dispersions for stellar systems
following Plummer density distributions with masses in the range of
$10^4 M_\odot$ to $10^9 M_\odot$ and which are either isolated or
immersed in an external field. Our integrations reproduce previous
analytic estimates for stellar velocities in systems in the deep
MOND regime ($a_i, a_e \ll a_0$), where the motion of stars is
either dominated by internal accelerations ($a_i \gg a_e$) or
constant external accelerations ($a_e \gg a_i$). In addition, we
derive for the first time analytic formulae for the line-of-sight
velocity dispersion in the intermediate regime ($a_i \sim a_e \sim
a_0$). This allows for a much improved comparison of MOND with
observed velocity dispersions of stellar systems. We finally derive
the velocity dispersion of the globular cluster Pal 14 as one of the
outer Milky Way halo globular clusters that have recently been
proposed as a differentiator between Newtonian and MONDian dynamics.
\end{abstract}
\begin{keywords}
galaxies: clusters: general- galaxies: dwarf - gravitation -
methods: analytical -- methods: N-body simulations
\end{keywords}
\section{Introduction}

The flattening of rotation curves of disk galaxies at large radial
distances, i.e. the apparently non-Newtonian motion, is usually
explained by invoking the otherwise undetected, so called Cold Dark
Matter (CDM) \cite{bos81,rub85}. This hypothesis has successfully
explained the internal dynamics of galaxy clusters, gravitational
lensing and the standard model of cosmology within the framework of
general relativity (GR). Despite the fact that the dark matter model
has been notably successful on large scales \cite{spe03}, dark
matter particles has not been detected after much experimental
efforts and the results of high resolution N-body simulations do not
seem to be compatible with observations on galactic scales
\cite{kly99,moore99,metz08}. Another approach to explain galaxy
rotation curves would be an alternative theory of gravity. One
promising alternative theory is Modified Gravity (MOG) which has
recently been successfully applied for dwarf satellite galaxies
\cite{mof07a} and distant globular clusters \cite{mof07b}. One of
the most famous alternative theories is the so-called modified
newtonian dynamics (MOND) theory, which was introduced by Milgrom
(1983). According to MOND, the flat rotation curves of spiral
galaxies at large distances can be explained by a modification of
Newton's second law of acceleration below a characteristic scale of
$a_{0}\simeq10^{-10} ms^{-2}$ without invoking dark
matter\cite{bek84}.

It has been shown that on galactic scales MOND can explain many
phenomena at least as well as CDM \cite{san02}. For example,
Sanchez-Salcedo and Hernandez (2007) studied the tidal radii of
distant globular clusters and dwarf spheroidal satellite galaxies in
MONDian dynamics. The most serious challenges for MOND come from
clusters of galaxies, where MOND cannot completely explain the
galaxy velocities \cite{san02}, and the merging of galaxy clusters,
where the baryonic mass is clearly separated from the gravitational
mass, as indicated by gravitational lensing \cite{Clo06}. Both
phenomena can be explained in MOND if some kind of hot dark matter
is assumed, perhaps in the form of a massive ($\sim 2$eV) neutrino
\cite{ang06}.

MOND has recently been generalized to a general-relativistic version
\cite{bek04}, making it possible to test its predictions for
gravitational lensing. In the non-relativistic version of MOND, the
acceleration of a particle due to a mass distribution $\rho$ is
given by \cite{bek84}:
\begin{equation}
{\bf \nabla}\cdot(\mu(\frac{a}{a_{0}}){\bf a})= 4\pi G
\rho=\nabla\cdot{\bf a}_{N}, \label{mond_pos}
\end{equation}
where ${\bf a}_N$ is the Newtonian acceleration vector, ${\bf a}$ is
the MONDian acceleration vector, $a=|\textbf{a}|$ is the absolute
value of MONDian acceleration and $\mu$ is an interpolating function
which runs smoothly from $\mu(x)=x$ at $x\ll1$ to $\mu(x)=1$ at
$x\gg1$. The standard interpolating function is
$\mu_1(x)=\frac{x}{\sqrt{1+x^2}}$, but Famaey \& Binney (2005)
suggested another function $\mu_2(x)=\frac{x}{{1+x}}$, which
provides a better fit to the rotation curve of the Milky Way.
Equation (\ref{mond_pos}) can be transformed into
$\nabla\cdot(\mu(\frac{a}{a_{0}}){\bf a}-{\bf a}_{N})=0$, where the
expression in parentheses is thus a curl field, and we may write
\begin{equation}
 \mu(\frac{a}{a_{0}}){\bf a} = {\bf a}_{N} + \nabla\times{\bf H}.\label{mon1}
\end{equation}
The value of the curl field ${\bf H}$ depends on the boundary
conditions and the mass distribution, but vanishes for some special
symmetries. 
In realistic geometries, the curl field is non-zero and leads to
difficulties for standard N-body codes. In other words, the
non-linearity of the MOND field equation makes the use of the usual
Newtonian N-body simulation codes impossible in the MOND regime.

Many stellar systems (e.g. globular clusters) have tidal radii much
larger than their sizes, therefore the external field is
approximately constant over the cluster area and the motions of
stars are not influenced by tidal effects.

In Newtonian dynamics, a stellar system evolving under the influence
of a uniform external acceleration, will, in the frame of the
system, have the same internal dynamics as an isolated system. In
MOND, due to the non-linearity of Poisson's equation, the strong
equivalence principle (SEP) is violated \cite{bek84}, and
consequently the internal properties and the morphology of a stellar
system are affected both by the internal and external field. This
so-called external field effect (EFE) significantly affects
non-isolated systems and can provide a strict test for MOND. The EFE
postulation originated from observations of open clusters in the
solar neighborhood, which do not show mass discrepancies even if the
internal accelerations are below $a_0$ \cite{mil83}. The EFE has
several consequences, for example it allows high velocity stars to
escape from the potential of the Milky Way \cite{fam07,wu07}, and it
decreases the velocity of satellite galaxies at very large radii,
which is in conflict with the asymptotically flattening of rotation
curves \cite{gen07,wu08}. The EFE implies that rotation curves of
spiral galaxies should fall where the internal acceleration becomes
equal to the external acceleration. In addition, if the EFE is taken
into account, internal properties of Galaxies such as the
Tully-Fisher relation should be changed \cite{wu07}.

Milgrom derived the mean velocity dispersion of stellar systems for
two special cases of internal or external field dominated systems

analytically, assuming that the systems are everywhere in the
deep-MOND regime ($a_e, a_i \ll a_0$). If the external acceleration
$a_e$ is much larger than the internal one $a_i$, the system of mass
$M$ is in the quasi Newtonian regime but with a normalized
gravitational constant larger than the standard Newtonian one by a
factor $\frac{a_0}{a_e}$, and therefore the line-of-sight velocity
dispersion is \cite{mil86}

\begin{equation}
 \sigma_{LOS,M1}=\sigma_{LOS,N}\sqrt{\frac{a_0}{a_e}},\label{milgrom86}
\end{equation}
where $\sigma_{LOS,N}$ is the Newtonian velocity dispersion. If
$a_e\ll a_i\ll a_0$, the cluster is isolated and the line-of-sight
velocity dispersion is given by

\begin{equation}
 \sigma_{LOS,M2}=0.471(GMa_0)^\frac{1}{4}.\label{milgrom94}
\end{equation}

Many systems which can be used to test MOND are not completely
internally or externally dominated, for example globular clusters or
dwarf galaxies of the Milky Way have internal and external
accelerations which are of the same order \cite{bau05}. Since
Milgrom's relations are valid only for systems that have either
$a_i\gg a_e$ or $a_i\ll a_e$ and are in the deep-MONDian regime, one
has to determine the velocity dispersions numerically for
intermediate cases. Milgrom found that for isolated systems
(internal acceleration dominated), the mass $M$ of a system is
nearly proportional to the forth power of the line of sight velocity
dispersion $\sigma_{los}$ and the ratio $\sigma_{los}^4/GM$ must be
somewhere between $\frac{4}{9}a_0$ and $a_0$. But how does the
velocity dispersion change while the system transits from the
Newtonian to the MONDian regime? In an attempt to answer this
question, we have performed N-body simulations and present
analytical formulae for the velocity dispersion of stellar systems
in the intermediate MOND regime. We have calculated the velocity
dispersion for a number of isolated systems in which the internal
accelerations $a_i$ are in the range from $a_i\ll a_0$ to $a_i\gg
a_0$. We also give formulae for systems with different strengths of
external fields. It should be noted that the isolated systems are in
equilibrium only in the Newtonian case, and reach a MONDian
equilibrium state after collapse. For non-isolated systems we start
from the MONDian equilibrium state which is created as described in
section 4.3. These results could be useful for comparison with
observational data of several GCs and dSph galaxies that are far
away from the host galaxy, so that the external acceleration due to
the host galaxy is small ($a_e<0.01 a_0$) these objects should
therefore provide straightforward possibilities to test MOND. Since
the external field affects the velocity dispersion by both tidal
effects and EFE, and in order to see the pure MONDian effects, we
concentrate on systems in which the tidal radius is much larger than
the virial radius and therefore tidal effects are unimportant.

This is the first of a series of papers that deals with the
numerical calculations for stellar systems. In the forthcoming
papers, the observational constraints on mass and velocity
dispersion of Pal 14 will be studied by Hilker et al. (2008) and
Jordi et al. (2008).

The paper is organized as follows: In Section 2 we introduce
theoretical predictions for the velocity dispersion in different
regimes. In Section 3, we give a brief review of the N-MODY code
which we use for our modelling. The numerical results for isolated
and non-isolated systems and comparison with observational data are
discussed in Section 4. We present our conclusion in Section 5.
\section{Line-of-sight velocity dispersion in different regimes}
\subsection{Newtonian regime}
In Newtonian gravity, the mean-square velocity, $\sigma^2$, of a
stellar system of mass $M$ is given by the following equation
(Equation (4-80a) of Binney and Tremaine (1987)):

\begin{equation}
 \sigma^2 = \frac{GM}{r_g},\label{dsip1}
\end{equation}
where $r_g$ is the gravitational radius defined as (Equation (2-132)
of Binney and Tremaine (1987)):

\begin{equation}
  r_g=\frac{GM^2}{|W|}. \label{rg}
\end{equation}
Here $W$ is the total potential energy. In the case of a Plummer
model \cite{plum}, and if we assume an isotropic velocity
distribution, the line-of-sight velocity dispersion becomes
\begin{equation}
\sigma_{LOS,N}=0.36\sqrt{\frac{GM}{R_h}}, \label{dis_p}
\end{equation}
where $R_h$ is the half-mass radius. If we define the
half-mass-radius acceleration as $a_h=\frac{GM}{2R_h^{2}}$ , we can
re-write the above relation as
\begin{equation}
\frac{\sigma_{LOS,N}^4}{GM}=0.035 a_h. \label{dis_n}
\end{equation}

\subsection{MOND regime}

In the case of MOND, and in the presence of an external field, the
total acceleration, which is the sum of the internal $a_i$ and
external $a_e$ acceleration, satisfies the modified Poisson equation
\cite{bek84},
\begin{equation}
 \nabla.[\mu(\frac{a_e+a_i}{a_{0}})({\bf a_i}+{\bf a_e})] \simeq 4\pi G \rho, \label{mon1}
\end{equation}
where $a_e$ is approximately constant, $a_i=\nabla\phi$ is the
non-external part of the potential and $\rho$ is the density of the
star cluster. The boundary condition is $\nabla\phi=a_e \hat{x}$ for
$r\rightarrow\infty$. Equation \ref{mon1} was postulated by Milgrom
(1983) to explain the dynamical properties of nearby open clusters
in the Milky Way and is an outcome of the MOND phenomenology. As an
approximation for a spherical system one can write equation
\ref{mon1} as:
\begin{equation}
a_i\mu(\frac{|a_e+a_i|}{a_{0}})=a_N \label{mon11}.
\end{equation}
Note however that Eq.\ref{mon11} is only an approximate and
effective way to take into account the external field effect (EFE),
in order to avoid solving the modified Poisson equation with an
external source term $\rho_{ext}$ on the right-hand side.
The EFE is indeed a phenomenological requirement of MOND, which has
important consequences for non-isolated systems. For example, if
$a_e \ll a_i\ll a_0$, then the dynamics is in the MOND regime, and
the external field can be neglected. When $a_i\ll a_e\ll a_0$, $\mu$
tends to its asymptotic value $\mu(a_e/a_0)=a_e/a_0$
(\emph{saturation of the $\mu$ function}), and the gravitational
potential is thus Newtonian with a renormalized gravitational
constant to ($G_{eff} = G/\mu(a_e/a_0)\approx Ga_0/a_e$
[\cite{mil86}]).\\
Recently, several papers were published using this
formulation to take into account the EFE
\cite{gen07,wu07,wu08,ang08,fam07,kly08}. For example, in order to
estimate the order of magnitude of the EFE, Famaey et al. (2007) and
Gentile et al. (2007), pointed out $a_0a(a+a_e)/(a_0 + a + a_e)=a_N$
using a simple $\mu$-function. Other authors replaced $|a_i+a_e|=
\sqrt{a_i^2+a_e^2}$ \cite{ang08, kly08}.  A more rigorous treatment
of EFE on galactic rotation curves was made by Wu et al. (2007,
2008). In the work by Wu et al. (2007), for the mass density of the
internal system, the MOND Poisson equation was solved as if the system
was isolated, but the boundary condition on the last
grid point was changed to be nonzero.\\
As a first approximation, we considered that a cluster is in a
non-inertial frame, which free-falls with a uniform systematic
acceleration. Since the calculation of $\phi$ is done for an
isolated cluster, we did not change the boundary condition and at
each step of potential solving, we added the constant external field
with $a_i$ inside the $\mu$ function. This method might be only an
approximation but as it is clear from the figures 1,3 and 4, the
transition region from Newtonian to MONDian case is reproduced
reasonably well by our method.

Analytical solutions exist only for some special cases that can be subdivided as follows:\\
1 - If $a_i\gg a_0$ or $a_e\gg a_0$ then the value of the
interpolating function is equal to one and the system is in the
Newtonian regime
and the velocity dispersion is given by equation (\ref{dis_n}).\\
2 - If $a_e\ll a_i\ll a_0$, the system is in the deep MOND regime
and the external field can be neglected (isolated system). In this
case, the line-of-sight velocity dispersion is given by Equation
(\ref{milgrom94}), which can be re-written as
\begin{equation}
\frac{\sigma_{LOS,M1}^4}{GM}=0.049a_0. \label{dis_m1}
\end{equation}
3 - If $a_i\ll a_e\ll a_0$, the system is externally field
dominated. $\mu(x)$ becomes in this case $\mu(a_e/a_0)=a_e/a_0=$
\emph{const} (saturation of the $\mu$ function) and the system goes
to a quasi-Newtonian regime but with an effective gravitational
constant $G_e=G\mu^{-1}(a_e/a_0)\simeq Ga_0/a_e$ that is larger than
the standard Newtonian one. Using equation (\ref{milgrom86}), the
line-of-sight velocity dispersion is therefore equal to,
\begin{equation}
\frac{\sigma_{LOS,M2}^4}{GM}= 0.035(\frac{a_0}{a_e})^2 a_h.
\label{dis_m2}
\end{equation}
Comparing equations (\ref{dis_n}), (\ref{dis_m1}) and (\ref{dis_m2})
with each other suggests that the line-of-sight velocity dispersion
in the general case should be given by
\begin{equation}
\frac{\sigma_{LOS}^4}{GM}=f(a_h), \label{dis}
\end{equation}
where $f(a_h)\sim a_h$ if $a_h\gg a_0$ and $f(a_h)\sim a_0$ if
$a_h\ll a_0$.

In this paper, we attempt to investigate the universal functional
form for $f(a_h)$ for systems with a wide range of internal and
external accelerations.
\section{N-MODY Code }
In order to numerically solve the non-linear MOND field equations,
recently two N-body codes have been developed \citep{cio06,tir07}.
In the present work we apply the N-MODY code developed by the first
group, which can be used to do numerical experiments in either
MONDian or Newtonian dynamics. N-MODY is a parallel,
three-dimensional particle-mesh code for the time-integration of
collision-less N-body systems \cite{lon08}. The potential solver of
N-MODY is based on a grid in spherical coordinates and is best
suited for modeling isolated systems. N-MODY uses the leap-frog
method to advance the particles. The code and the potential solver
have been presented and tested by Ciotti et al. (2006) and Nipoti et
al. (2007).

In the present study we used a spherical grid ($r,\theta,\varphi$)
made of $N_r\times N_\theta \times N_\varphi=64\times64\times128$
grid cells for the integration. We use twice as many cells in the
$\varphi$ direction since $\varphi$ runs from $0<\varphi<2\pi$ while
$\theta$ runs only from $0<\theta<\pi$.
The total number of particles was in the range $N_p = 10^5 - 10^6$.
The details of the scaling of the numerical MOND models and code
units are discussed in Nipoti, Londrillo \& Ciotti (2007). In order
to include the EFE for non-isolated systems, we changed the N-MODY
code and put a constant external field in the MONDian potential
solver. We also chose $\textbf{a}= \textbf{a}_i+\textbf{a}_e$ within
the interpolating function as the total acceleration of particles.

In the present work, the Plummer model (Plummer 1911) was used as
the initial cluster model.  It has a density distribution
\begin{equation}
\rho(r) = \frac{3M}{4 \pi
r_{Pl}^3}\left(1+\frac{r^2}{r_{Pl}^2}\right)^{-5/2} \label{plumer}
\end{equation}
where $M$ is the total mass and $r_{Pl}$ is the 'scale radius'. The
half-mass radius of a Plummer model is $R_h \simeq 1.305 r_{Pl}$ and
the virial radius is $R_v= \frac{16}{3 \pi}r_{Pl}$. The total
potential energy, $|W|=\frac{3\pi}{32}\frac{GM^2}{r_{Pl}}$, is used
in equation (\ref{rg}) to calculate $r_g$.

\begin{figure}{}
\begin{center}
\resizebox{9.3cm}{!}{\includegraphics{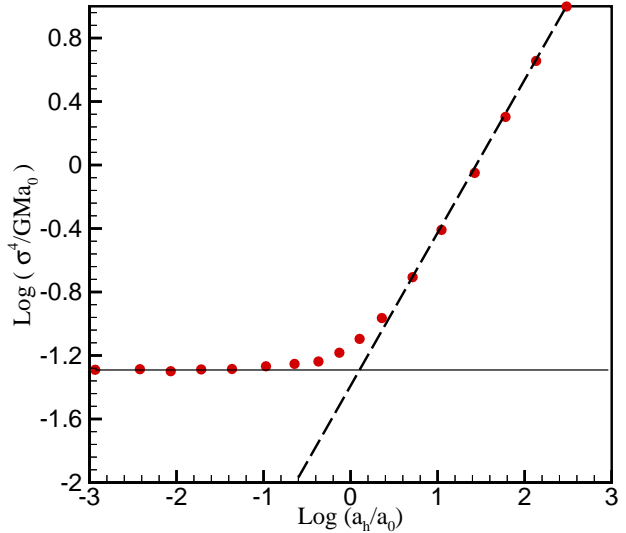}} \caption{
Line-of-sight (LOS) velocity dispersion profiles for isolated
stellar systems as calculated by N-MODY. In order to have different
internal half-mass accelerations $a_h$, several cases with different
half-mass radii were calculated. As expected from equation
(\ref{dis}), all curves follow the same functional form. The dashed
line shows the asymptotic behavior in the Newtonian (equation
\ref{dis_n}) regime. The solid line shows the asymptotic behavior in
deep MONDian (equation (\ref{dis_m1})) regime. For high internal
acceleration ($a_h\gg a_0$), the models are consistent with the
Newtonian result and for low acceleration ($a_h\ll a_0$), they are
consistent with the deep MONDian prediction. ($Log\equiv log_{10}$).
\label{fig1}}
\end{center}
\end{figure}

\begin{figure}{}
\begin{center}
\resizebox{9.3cm}{!}{\includegraphics{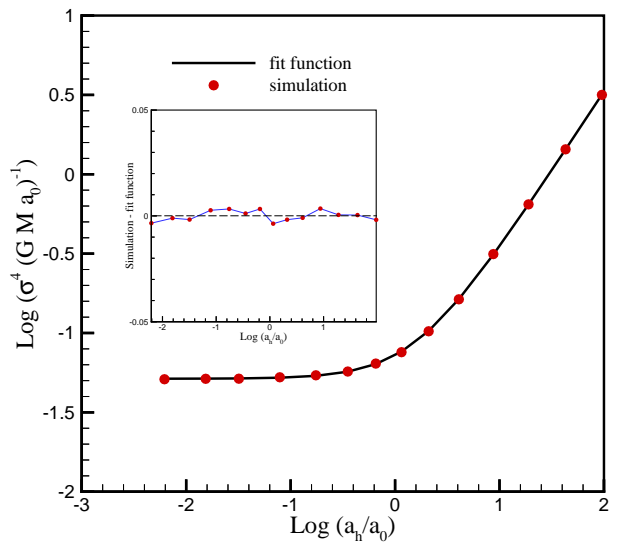}} \caption{Fit of our
best fitting curve (equation (\ref{fitfunc})) to the numerical
solution for an isolated stellar system. The difference of each
point from the fit function is presented in the inset. The average
residual of this function from our numerical solution, which is
defined as $\Delta=|f_{theory}-f_{fit}|$, is less then $10^{-3}$.
($Log\equiv log_{10}$). \label{fit_iso}}
\end{center}
\end{figure}

\begin{figure}{}
\begin{center}
\resizebox{9.3cm}{!}{\includegraphics{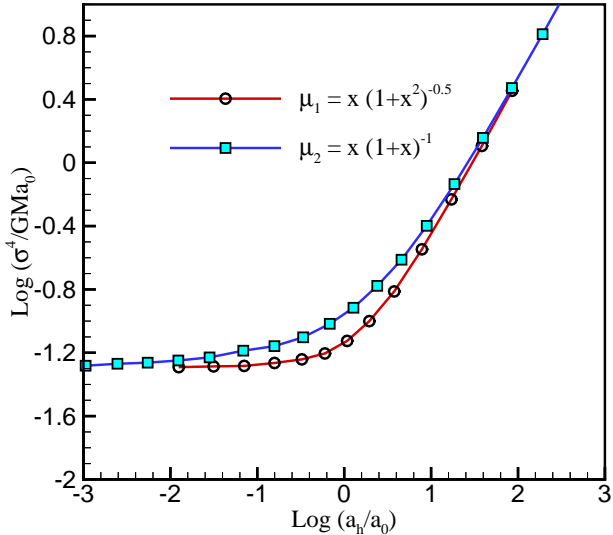}} \caption{Effect of
different choice of interpolation function on the line-of-sight
velocity dispersion for systems with different internal
accelerations. Both functions have the same value in the Newtonian
and the deep MONDian regime. In the transition zone, the simple
function, $\mu_2$, produces a larger velocity dispersion. This means
that if the observed velocity dispersion of a stellar system shows a
value smaller than the MONDian prediction with the standard
interpolation function, $\mu_1$, the simple function $\mu_2$ would
not help to decrease this discrepancy. The largest difference
between both functions is of order $20\%$ and occurs at $a_h=a_0$.
($Log\equiv log_{10}$). \label{mu}}
\end{center}
\end{figure}

\begin{figure}{}
\begin{center}
\resizebox{9.3cm}{!}{\includegraphics{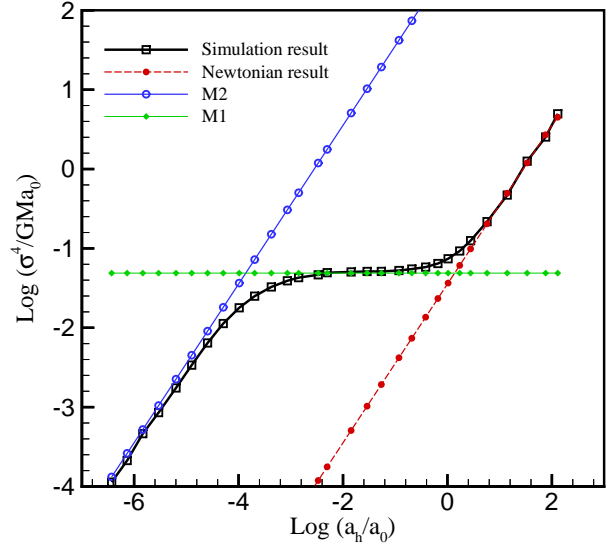}} \caption{ The
line-of-sight velocity dispersions for stellar systems with an
external field of $a_e=0.01 a_0$ with different internal
half-mass-radii accelerations as calculated by N-MODY (black line
with open squares). $M1$ refers to equation (\ref{dis_m1}) which is
for the isolated system and $M2$ refers to the quasi Newtonian case
dominated by the external field (equation (\ref{dis_m2})). Due to
the saturation of the $\mu$ function in the external field dominated
regime, the velocity dispersion curve (open squares) starts to fall
in a quasi Newtonian way (blue line with open circles) with
decreasing $a_h$ and deviates from the prediction of MOND for
isolated systems, $M1$, (green solid line with closed diamond).
Moving towards decreasing $a_h$, the first transition occurs near
$a_h\approx a_0$, when the system enters into the MONDian regime and
the velocity dispersion deviates from the Newtonian prediction (red
dashed line with filled circles). The horizontal axis gives the
Newtonian half-mass-radius acceleration. Since the Newtonian
internal acceleration of a system is the square of the MONDian
acceleration ($a_M=\sqrt{a_Na_0}$), the point
$log_{10}(\frac{a_{h,N}}{a_0})=-4$, corresponds to $a_{h,M}=0.01
a_0$ in MOND. The velocity dispersion remains on the horizontal line
which corresponds to the isolated system until the internal
acceleration reaches $a_i \approx a_e \approx 0.01 a_0$. ($Log\equiv
log_{10}$). \label{fig4}}
\end{center}
\end{figure}

\begin{figure}{}
\begin{center}
\resizebox{9.3cm}{!}{\includegraphics{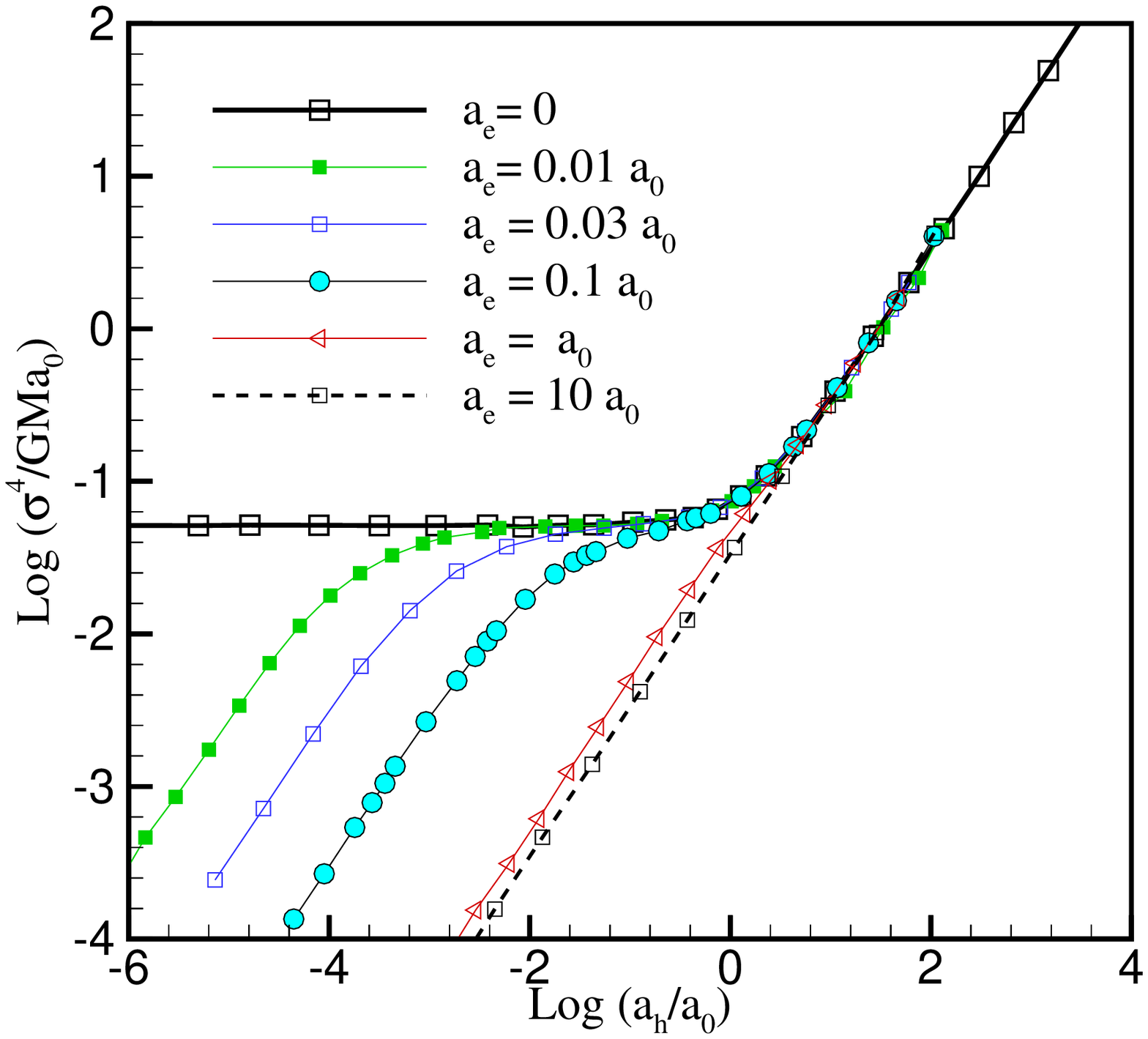}} \caption{External
field effect on predicted line-of-sight velocity dispersions for
stellar systems with different internal accelerations as calculated
by N-MODY. The x-axis gives the Newtonian internal acceleration of
the system. In order to see the transition regime we assume several
values of $a_e$. When the internal acceleration of the system
decreases, there are two transitions in the velocity treatment. The
first transition occurs near $a_h\approx a_0$ from Newtonian into
MONDian regime and the second transition from the MONDian to quasi
Newtonian regime occurs when the internal acceleration becomes equal
to the external acceleration. The functional form of each fit curve
is given in Table (\ref{table}). ($Log\equiv log_{10}$). \label{ae}}
\end{center}
\end{figure}

\section{ Results }
\label{S5} In this Section, we present N-MODY solutions for stellar
systems that are both isolated and non-isolated, allowing for
different values of the external field.

\subsection{Isolated systems}

We have performed a large set of dissipationless N-MODY computations
for isolated systems. Since the modeled systems are in equilibrium
in the Newtonian case, in the MONDian case, they initially collapse.
In order to have a MONDian equilibrium initial system, we rescaled
the velocities by an amount given by our fitting formulae ( Section
4.2 Equations 15 ) to prevent collapse. In order to create the
initial condition, there is another substantial method developed by
Nipoti et al. (2007b, 2008), in which the distribution function is
obtained numerically with an Eddington inversion with the far field
logarithmic behavior of the MOND potential. Here we have used our
method to set up MONDian initial condition. This method could
generalis to non isolated systems easily (see section 4.2).

\begin{figure*}
\centering
\includegraphics[width=168mm]{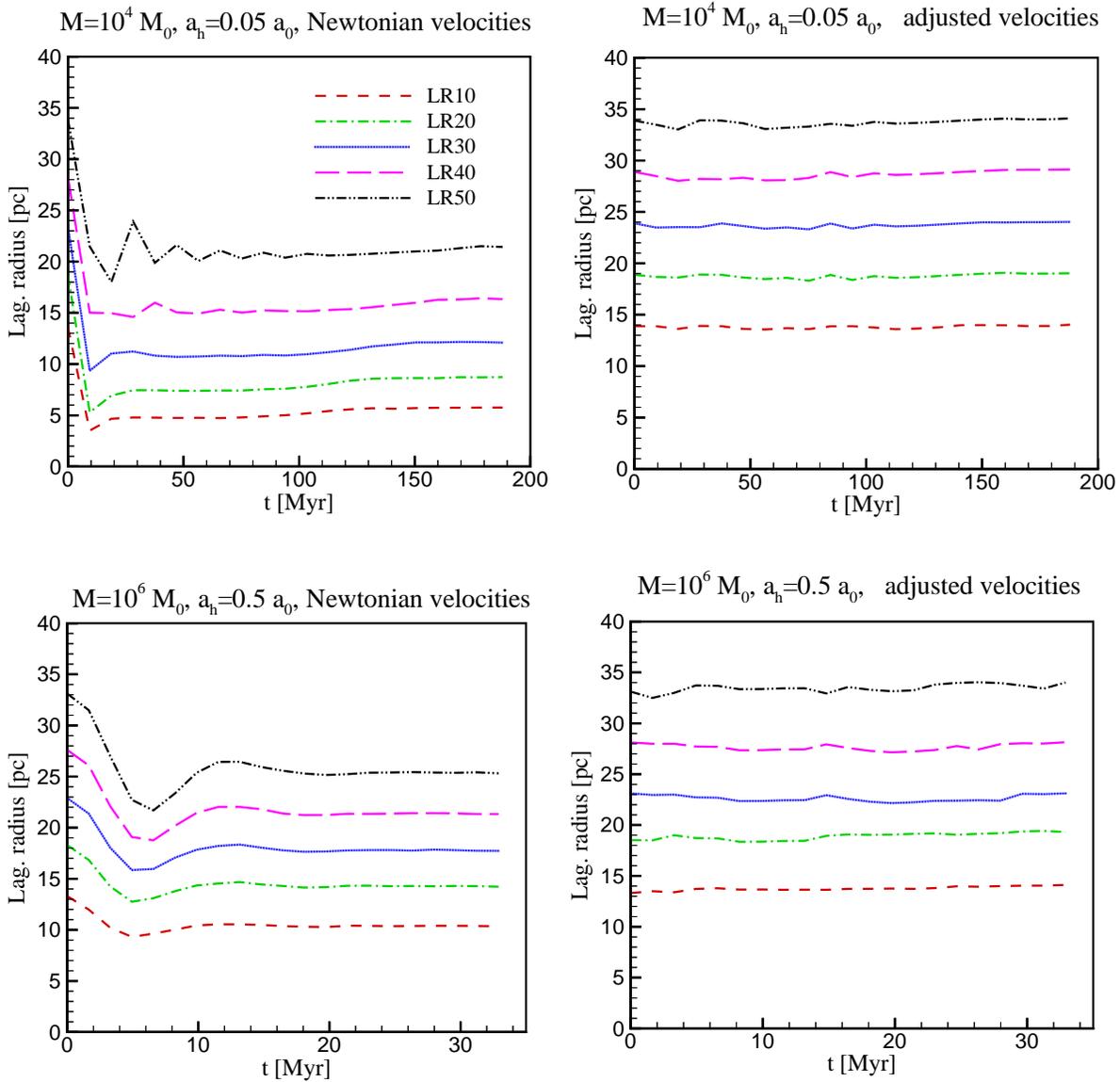}
\caption{Upper left: Evolution of Lagrangian radii for a low mass
cluster with stellar velocities corresponding to Newtonian virial
equilibrium but in the deep-MOND limit. After rapid collapse, the
system reaches an equilibrium state. Upper right: By increasing the
initial velocities of particles by a factor of 2.8, calculated from
equation (15), the system reaches equilibrium in MOND without
collapsing. Lower panel: Evolution of the Lagrangian radius for a
high mass cluster in the intermediate MOND regime. In the left
panel, the velocities are not adjusted and the system still
collapses. After adjusting the velocities by a factor of 1.38,
calculated from equation (15), there is no collapse (right panel).
}\label{lag}
\end{figure*}

As discussed in Nipoti et al. (2007a), all simulations with
different masses but with the same value of $a_h$ are identical, in
the sense that they can be simply rescaled to different masses,
provided that $M/r_h^2 = 2a_h/G$ remains constant. As a consequence,
systems of any mass with $a_h$ in the explored range follow the same
functional form. Therefore, we consider only one simulation for
given $a_h$. In order to produce different internal acceleration
regimes, we changed the half-mass radii of the system from 1 pc to 1
kpc. The models are evolved for several crossing times to reach the
equilibrium state, which is identified by stationary Lagrange radii
(e.g. Fig \ref{lag}).

The resulting global velocity dispersions as a function of internal
acceleration of the stellar systems are plotted in Fig. \ref{fig1}.
As expected from equation (\ref{dis}), all of them follow the same
functional form. The dashed line shows the Newtonian prediction for
the velocity dispersion (equation (\ref{dis_n})). The asymptotic
behavior of the models in the Newtonian regime are compatible with
this analytical prediction. The solid line shows the analytical
velocity dispersion in the deep MONDian regime (equation
(\ref{dis_m1})). In the low acceleration region, the numerical
solutions are compatible with the analytical formula. At $a_h =
a_0$, the difference between the numerical model and the MONDian
prediction is about $0.2$ in $log_{10}$, which means that
$\sigma(a_h)\simeq 1.3\times\sigma_{LOS,M1}$.

We now try to find an expression for a function $f_0(x)$ where
$x=\frac{a_h}{a_0}$ which fits the numerical results. In the
Newtonian regime $(x\gg1)$, the function $f_0$ has to approach
$f_0(x)=x+const$, while in the deep MONDian regime $(x\ll1)$, $f_0$
has to be constant. We therefore make an ansatz,
\begin{equation}
f_0(x)=a\ln(\exp(\frac{x}{a})+b)+c, \label{fitfunc}
\end{equation}
for the function $f_0$. The best-fitting coefficients are then
determined by a least-squares fit to the data and are found to be
$a= 0.3314$, $b= 1.78$, and $c= -1.48$. This function is shown in
Fig. \ref{fit_iso} as a solid line. The average residual of this
function from our numerical results, which is defined as
$\Delta=|f_{theory}-f_{fit}|$, is less then $10^{-3}$. Therefore,
for any isolated system, if the internal half-mass-radius
acceleration, $a_h$, can be measured, it is possible to find out the
MONDian prediction by this function. This is especially useful for
the intermediate case which has no analytical prediction in MOND.
The corresponding formula for the velocity dispersion is

\begin{eqnarray}
\log_{10}(\sigma_{LOS})&=&0.25\{0.331\ln[\exp\left(\frac{1.51GM}{a_0R_h^2}\right)
+1.78]\\&&-1.48+\log_{10}(GMa_0)\}. \label{dissim}\nonumber
\end{eqnarray}
A simple relation exists between the three-dimensional half-mass
radius and easier to observe two-dimensional, projected half-mass
radius $R_{hp}:R_{hp}=\gamma R_h$ with $\gamma \approx 0.74$, which
can be used in this formulae.

We briefly also discuss the influence of a different $\mu$ function
on the results. In Fig. \ref{mu} we plot the velocity dispersion for
an isolated system using the simple interpolation function $\mu_2$,
and compare it with the results obtained for $\mu_1$. Since the
simple function has a stronger MONDian effect, the velocity
dispersion is higher than that of the standard function. The
difference of the velocity dispersion between both functions at
$a_h=a_0$ is about 20\%, so in order to determine the $\mu$ function
from observations, one needs to measure the mass and overall
structure of a stellar system very accurately. As expected, in the
extreme limit of $a_h \ll a_0$ or $a_h \gg a_0$, both functions
predict the same value for the velocity dispersion.

\subsection{Non isolated systems}

Systems relevant for testing MOND (e.g. globular clusters or dwarf
galaxies) usually move through the gravitational field of a host
galaxy. Therefore, the internal dynamics is often influenced by the
host galaxy due to the EFE of MOND. We assume that coriolis forces
that arise in the rotating reference frame of the cluster and that
tidal forces arising from a gradient of the external field can be
neglected. We believe this to be a a good first approximation that
allows us to focuss on the effects of the (constant) external field,
therewith allowing us to for the first time venture into the
intermediate MOND regime in order to study the external field effect
numerically.

As an example which shows the EFE on the predicted LOS velocity
dispersion for non-isolated stellar systems with different internal
accelerations, we choose an external acceleration of $a_e=0.01 a_0$.
This corresponds to a cluster or dwarf galaxy being at a distance of
about $1Mpc$ from the Galactic center for an enclosed Milky Way mass
of $10^{12} M_{\odot}$. The resulting velocity dispersion as
calculated by N-MODY is shown in Fig. \ref{fig4}. The first
transition occurs near $a_h\approx a_0$, when the systems enter the
MONDian regime and the velocity dispersion deviates from the
Newtonian prediction. The velocity dispersion remains close to the
MOND prediction for the isolated case until the internal
acceleration reaches $a_i \approx a_e \approx 0.01 a_0$ at which
point a second transition occurs. Due to the saturation of the $\mu$
function in the external field dominated regime, the velocity
dispersion curve falls in a quasi Newtonian way if  $a_h < a_e$ and
therefore deviates from the prediction of MOND for isolated systems.

Note that the internal acceleration shown in Fig. \ref{fig4} is the
Newtonian internal acceleration, $a_h=\frac{GM}{2r_h^2}$, of the
system. In the deep MOND regime, since the Newtonian acceleration is
the square of the MONDian acceleration ($a_M=\sqrt{a_Na_0}$), the
point $a_{h,N}=0.0001 a_0$ corresponds to $a_{h,M}=0.01 a_0$.

In order to see the effect of different external accelerations, the
MONDian velocity dispersion as a function of internal acceleration
is plotted in Fig. \ref{ae} from a weak to a strong external field.
Note that the transition point is determined by the strength of the
external field. For a smaller external field, the transition point
occurs at a smaller acceleration. As predicted by theory, for a
strong external field ($a_e\gg a_0$) the system is completely in the
Newtonian regime, even for a low internal acceleration.

In order to find out the best functional form for the velocity
dispersion as a function of the strength of the external field, we
use the same procedure as in the isolated case, but take into
account the different asymptotic behavior in Fig. \ref{ae}. In the
Newtonian regime, $(x\gg1)$, $f(x)$ still changes as $f(x)=x+const$.
In the deep MONDian regime $(x\ll1)$, the systems are in the
quasi-Newtonian regime and $f(x)$ is again linear with the same
slope as in the Newtonian regime. In the intermediate regime, in
which the system is internal-acceleration dominated,
$\sigma_{los}^4/GM$ has to be constant. A general function
satisfying all these constraints is therefore given by
\begin{equation}
f(x)= f_0(x)- a\ln(\exp(-\frac{x}{a})+b)+c. \label{fitfunc-ae}
\end{equation}
The coefficients $a$, $b$ and $c$ depend on the external
acceleration. Table \ref{table} gives their values for several
values of the external acceleration $a_e$. Since the asymptotic
value of different interpolating functions is the same, the choice
of the $\mu$-function does not affect systems which are in the low
acceleration regime. However for the higher acceleration systems
($a_h\sim a_0$), the $\mu$-function plays a more important role.
Equation (\ref{fitfunc-ae}) will for example allow it to test MOND
against the observed global velocity dispersions of dwarf galaxies.
The relevant external acceleration can be determined by
interpolating between the points computed with N-MODY (Table
\ref{table}).

\begin{figure}{}
\begin{center}
\resizebox{9.3cm}{!}{\includegraphics{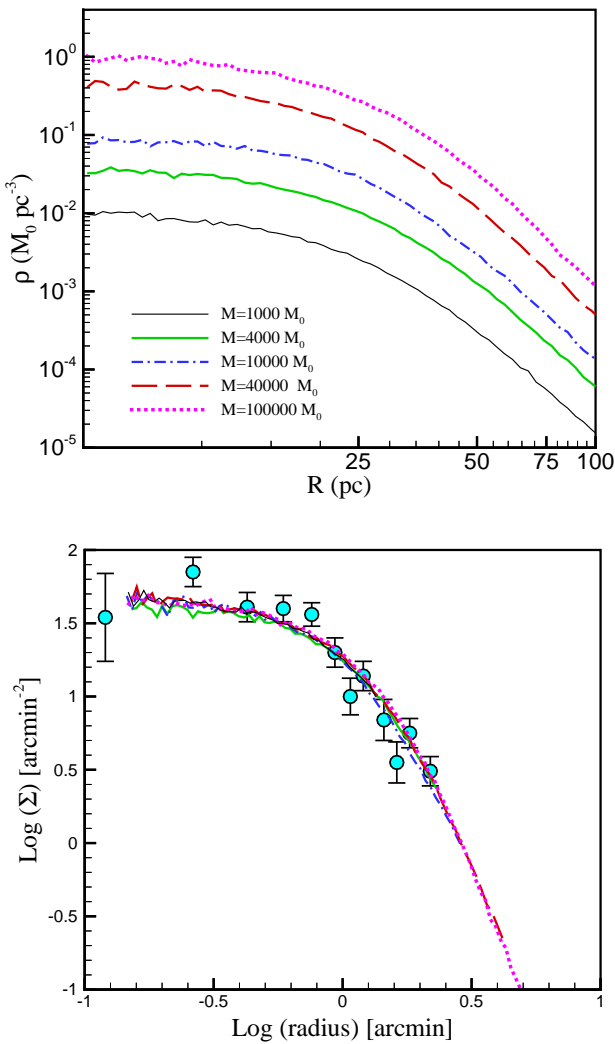}} \caption{Upper
panel: Density profile of Pal 14 for different cluster masses
obtained by N-MODY computations. The shapes of the profiles are the
same for all masses. Lower panel: Surface density profile of Pal 14
for masses as in the upper panel scaled to the level of data. The
surface density profile shapes compare well with the observed
density profile of Pal 14 (blue dots) as traced by giant stars
(Hilker 2006). The meaning of the different lines is as in the upper
panel. ($Log\equiv log_{10}$). \label{dens}}
\end{center}
\end{figure}

\begin{figure}{}
\begin{center}
\resizebox{9.1cm}{!}{\includegraphics{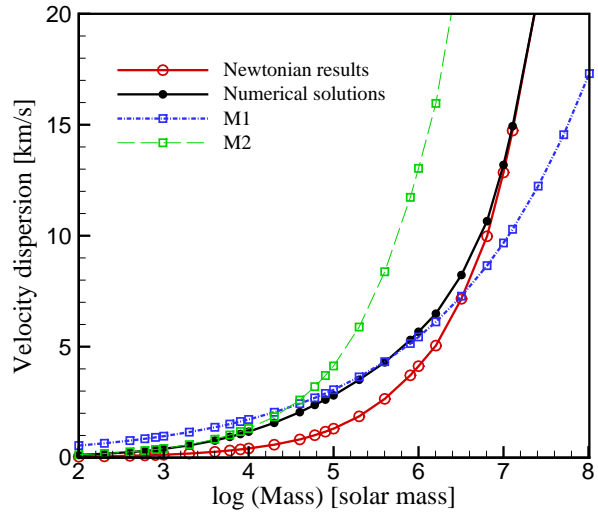}} \caption{
Line-of-sight velocity dispersion for Pal 14 for various masses as
found by N-MODY. In order to compare with the real cluster
(observational velocity dispersion of Pal 14), the half mass radii
of all models are fixed at 33 pc. The analytical predictions for
different limiting cases are also plotted for comparison. $M1$
refers to equation (\ref{dis_m1}), (isolated systems) and $M2$
refers to the quasi-Newtonian case (external-field dominated case).
For low masses, which means low internal accelerations, the
prediction of $M2$ is compatible with the numerical solution. As the
mass increases, the internal acceleration grows and the system
enters the internally-dominated regime and the numerical solutions
get close to the $M1$ prediction. \label{paldip}}
\end{center}
\end{figure}

\subsection{The velocity dispersion of Galactic globular clusters}

In order to decide whether MOND or dark matter is the right theory
to explain the dynamics of the universe, it is desirable to study
MOND for objects in which no dark matter is supposed to exist and
where the characteristic acceleration of the stars is less than the
MOND critical acceleration parameter $a_0$. GCs are a perfect
candidate since they are the largest virialized structure that does
not contain dark matter \cite{moore96}, and their internal
accelerations can be lower than $a_0$. Hence, GCs may provide a good
laboratory to test the law of gravity \cite{bau05}.

We choose the globular cluster Pal 14, for which there is a current
observational effort to determine its velocity dispersion (Jordi et
al. 2008). We initially choose a Newtonian equilibrium Plummer model
initially. While the half-mass radius of Pal 14 is about 33 pc
(Hilker 2006), the mass is not actually known, but an observing
campaign is underway to constrain it (Hilker et al. 2008). We change
the mass in the wide range from $[10^2 - 10^7] M_{\odot}$ and
consider the half-mass radius to be constant. We perform numerical
modeling to obtain the mean velocity dispersion as well as the
density profile and velocity dispersion profile.

Since the modeled systems are in equilibrium in the Newtonian case,
in the MONDian case, they initially collapse and $R_h$ is decreased
before the systems virialize again. In order to have a MONDian
equilibrium initial system, we increased the velocity of the
Newtonian system by an amount given by our fitting formulae
(Equations (15) and (\ref{fitfunc-ae})) to avoid a collapse. For low
mass systems that are in the deep-MOND regime ($a_i\ll a_0$) the
increase is larger than for massive systems that are in the
intermediate regime ($a_i\sim a_0$). In Fig. \ref{lag} we plot the
evolution of the Lagrangian radii for two clusters with the same
half-mass radius and different mass in the deep-MOND and
intermediate MOND regime. In deep-MOND (low mass cluster), after a
rapid collapse, the system reaches an equilibrium state. By
increasing the initial velocity of the particles, the collapse can
be prevented.

In Figs. \ref{dens} and \ref{paldip} we show the numerical solution
for Pal 14. We plot the density profile of Pal 14 for different
masses and compare it with the observed profile in Fig. \ref{dens}
(observational data from Hilker (2006)). The shape of the density
profiles is the same for all masses, but the central density differs
significantly. All calculated surface density profiles compare well
with the observed density profile. It should be mentioned that the
full observed density profile is not known for Pal 14 and that the
surface density profile shown in Fig. \ref{dens} is based only on
giant stars.

Fig. \ref{paldip} shows the numerical solutions for the
line-of-sight velocity dispersion  and compares it with Milgrom's
analytical predictions for the extreme limits (to see how analytical
predictions differ from the numerical solution). $M1$ refers to
equation (\ref{dis_m1}) which is for isolated systems and $M2$
refers to the quasi-Newtonian case which is for the external field
dominated case (equation (\ref{dis_m2})). As expected, the
analytical estimates are consistent with the numerical solution
either in the external field dominated (small mass) or internal
field dominated case, but have significant deviations in the
intermediate regime. For Pal 14 the external acceleration of the
Galaxy is about $a_e \sim 0.1 a_0$. For low cluster masses which
means low internal accelerations, the prediction of $M2$ is
compatible with the numerical solution. As the mass increases, the
internal acceleration grows and the system enters the internally
dominated regime and the numerical solution gets close to the $M1$
prediction.

\begin{figure*}
\centering
\includegraphics[width=168mm]{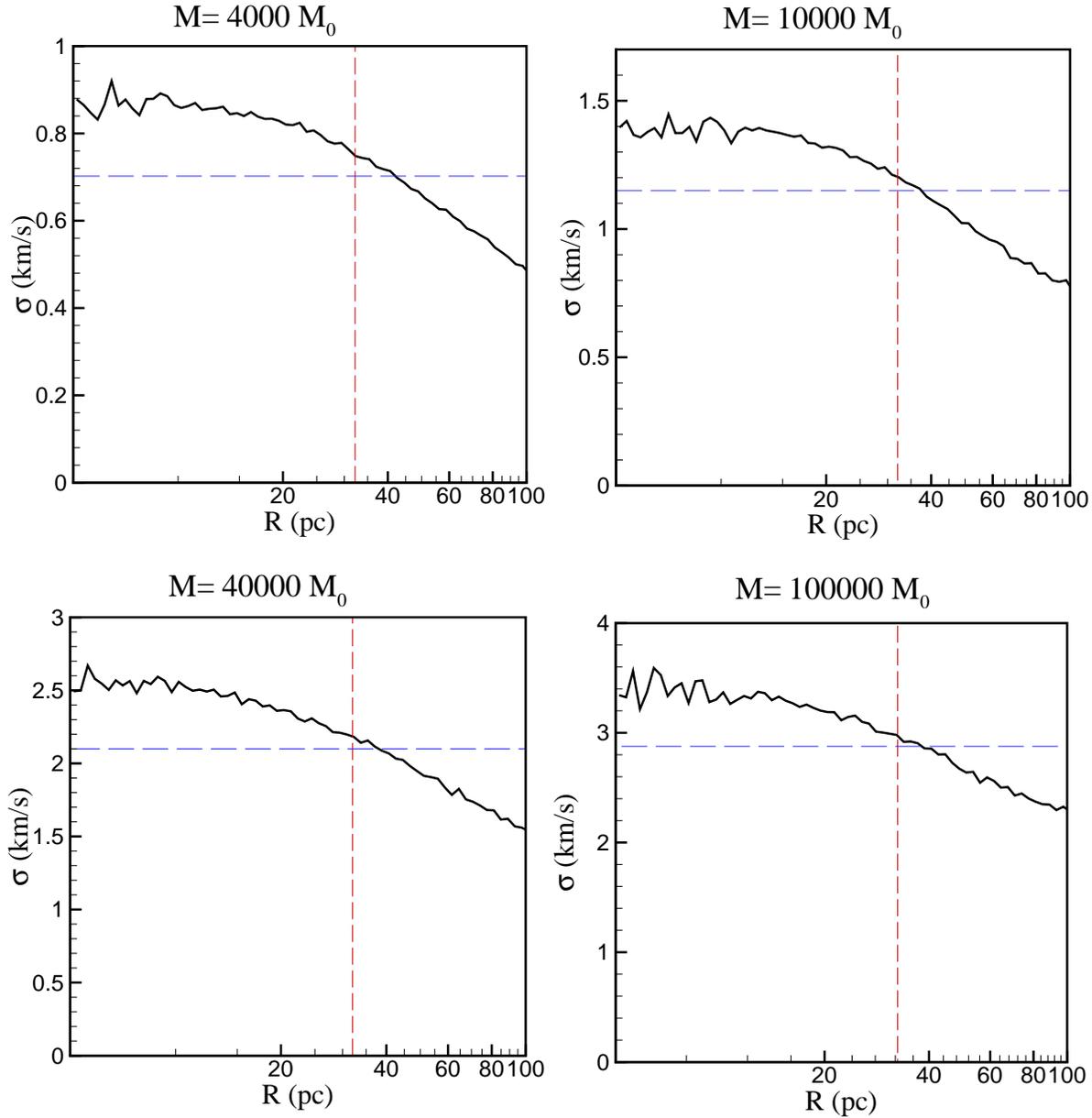}
\caption{Line-of-sight velocity dispersion profiles of Pal 14 for
different cluster masses obtained by N-MODY computation. The
horizontal blue (dashed) line indicates the mean global velocity
dispersion and the vertical red (dashed) line indicates the
half-mass-radius of Pal 14. The velocity dispersion changes only
slightly (about $10\%$) from the center to the half-mass-radius.
}\label{disprof}
\end{figure*}

The velocity dispersion profiles of Pal 14 for different cluster
masses are also plotted in Fig. \ref{disprof}. The velocity
dispersion changes slightly (about $10\%$) from the center to the
half-mass-radius of the cluster.

We would finally like to mention that the line-of-sight velocity
dispersion in the direction of the external field is almost $5\%$
lower than that perpendicular to the external field. Such a small
difference would be unobservable. This can be understood as arising
from the external field negating MONDian gravity in the direction of
the field. Also, the cluster is elongated by a few percent along the
vector $\textbf{R}_g$. However for larger systems such as galaxies,
the anisotropy could be observable for a case in which $a_i \ll a_e
\ll a_0$.

\begin{table}
\begin{center}
\begin{tabular}{|c|c|c|c|c|c|}
\hline
External acceleration          & $a$ & $b$ & $c$    \\
\hline      $a_e=0.01a_0$  &0.489 &2242&-7.694  \\
\hline      $a_e=0.03a_0$  &0.496&292&-5.680                  \\
\hline      $a_e=0.1a_0 $  &0.495&35.32& -3.585   \\
\hline      $a_e=0.3a_0 $  &0.342&8.38&-2.119            \\
\hline      $a_e=1.0a_0 $  &0.281&1.00& -0.006 \\
\hline      $a_e=10.0a_0$  &0.378&0.56&0.477 \\
\hline
\end{tabular}
\end{center}
\caption{Best fitting coefficients for the velocity dispersion
predicted by N-MODY simulation for various values of the external
acceleration. The general form is given by equation
(\ref{fitfunc-ae}) , where $f_0(x)$ is due to an isolated cluster
(equation (\ref{fitfunc})). In case $a_e=10 a_0$, the function is
nearly linear and the best fit function can also be obtained by
$f(x)=x-1.467$, i.e. the Newtonian prediction. \label{table}
 }
\end{table}

\section{Conclusions}

In this work we have calculated global line-of-sight velocity
dispersions of stellar systems in MOND for both isolated and
non-isolated stellar systems. The velocity dispersion of stellar
systems in MOND was so far only known in the case of the deep
MONDian limit where all accelerations are much smaller than the
critical acceleration, $a_0$, and even in this case only if either
the internal acceleration is much larger than the external
acceleration or the internal acceleration is much lower than the
external acceleration. We used the N-MODY code to calculate for the
first time the line-of-sight velocity dispersions of stellar systems
also for the intermediate regime.

We have obtained a large set of dissipationless N-MODY numerical
solutions for isolated systems with masses in the range $10^{4}
M_\odot$ to $10^{9} M_\odot$ and with the Plummer model as the
initial condition. In order to produce different internal
acceleration regimes, for each mass, we changed the half-mass radius
of the system. We deduce the analytical formulae for the velocity
dispersion of a stellar system as a function of its
half-mass-radius-internal-acceleration, $a_h$ (equation
(\ref{dis})), and investigate the universal functional form for the
velocity dispersion of isolated systems (equation (\ref{fitfunc})).

We have also studied the effect of a different choice of the
interpolation function on the line-of-sight velocity dispersion for
systems with different internal accelerations. We found that the
simple function suggested by Famaey and Binney (2005) produces a
larger velocity dispersion than the prediction with the standard
interpolation function suggested by Milgrom (1984), with the maximum
difference occurring at $a_h\sim a_0$ and being of order $20\%$.

Since most stellar systems (e.g. globular clusters or dwarf
galaxies) are not isolated and usually move through the
gravitational field of a host galaxy, the internal dynamics is often
influenced by the host galaxy due to the external field effect (EFE)
of MOND. Therefore, we have investigated non-isolated systems,
adding the external field to N-MODY. Our simulations reproduce
previous analytic estimates for stellar velocities in systems in the
deep MOND regime ($a_i, a_e \ll a_0$), where the motion of the stars
is either dominated by internal accelerations ($a_i \gg a_e$) or
external accelerations ($a_e \gg a_i$). In addition, we calculate
the line-of-sight velocity dispersion for intermediate cases and
derive for the first time analytic formulae for the line-of-sight
velocity dispersion in the intermediate regime ($a_i \sim a_e \sim
a_0$) and found a smooth functional form for the velocity dispersion
of stellar systems under the EFE. These formulae will allow to test
MOND more thoroughly than was hitherto possible.

We finally calculated the velocity dispersion of the globular
cluster Pal 14, and will compare it with observational data in a
forthcoming paper (Jordi et al. 2008). An additional observational
study in order to constrain the mass of Pal 14 is also underway by
our team (Hilker et al. 2008). In a future contribution we will also
discuss the fascinating possibility of "freezing" a cluster on a
highly eccentric orbit: as a cluster moves from the Newtonian regime
(small $\sigma_N$) to the MONDian regime on a time scale comparable
to or faster than the internal crossing time it will retain a
Newtonian velocity dispersion (Haghi et al. 2008).

Additional observational efforts to determine the velocity
dispersion of stellar systems such as GCs or dSph satellites would
be highly important as such data also provide a strict test of MOND.
On the other hand, if we believe in MOND, these observations could
be used to constrain the external field and consequently to put
constrains on the potential in which the systems are embedded.
Moreover it would be worthwhile to observe the stellar system in the
intermediate regime to constrain the $\mu$-function and $a_0$.
\label{S6}
\section*{Acknowledgements}
We would like to thank C. Nipoti for providing us with the N-MODY
code and his help in using it. H.H thanks the Iranian Cosmology and
Particle Physics Center of Excellence, at the physics department of
Sharif University of Technology and the Argelander Institute for
Astronomy for providing fellowships in support of this research. K.J
and E.K.G gratefully acknowledge support by the Swiss National
Science Foundation.

\end{document}